\documentstyle[prb,aps,psfig]{revtex}
\begin{document}

\title{Superconductor--insulator transitions and insulators with localized
pairs}

\author{V.Gantmakher \thanks{e-mail: gantm@issp.ac.ru}}

\address{Institute of Solid State Physics, Russian Academy of
Sciences, 142432 Chernogolovka, Russia}
\maketitle

\begin{abstract}
Two experiments are described which are related to the problem of localized
Cooper pairs. Magnetic-field-tuned superconductor-insulator transition was
studied in amorphous In--O films with onset of the superconducting transition
in zero field near 2~K. Experiments performed in the temperature range
$T>0.3\,K$ indicate that at the critical field, $B=B_c$, the first derivative
of the resistance $\partial R/\partial T$ is non-zero at $T=0$ and hence the
scaling relations should be written in more general form. Study of the
magnetotransport of high-resistance metastable alloy Cd--Sb on the insulating
side of the superconductor-insulator transition revealed below 0.1~K a
shunting condiction mechanism in addition to usual one-particle hopping.
Possibility of pair hopping is discussed.

\end{abstract}

Among various scenarios of superconductor--insulator transitions (SIT) there
exists a specific one proposed by Fisher \cite{Fisher}: a field-induced and
field-tuned transition in two-dimensional superconductors. It supposes, at
$T=0$, existence of delocalized Cooper pairs and localized vortices
(superconductor) below the transition, at fields $B<B_c$, and of localized
pairs and delocalized vortices (insulator) above it, at $B>B_c$. Several
experimental studies \cite{HebPaa,Kapit} apparently support this model
describing the experimental data near the transition in terms of scaling
relations \cite{Fisher}. As a consequence of the approach of
\cite{Fisher,HebPaa,Kapit}, an insulator should be supposed to exist with
localized Cooper pairs and with structurized into vortices magnetic field.
Properties of such insulator still have not been addressed. However, some
experimental observations do point to localized pairs \cite{HP}. In
particular, negative magnetoresistance in some three-dimensional materials,
when superconductivity is alternated by insulating behavior, is one of such
indications \cite {ShOv}. This negative magnetoresistance results from
destroying by the magnetic field of the gap in the spectrum of localized
electrons, which affects the single-particle tunneling \cite{gg}.

Still, our knowledge of localized pairs is rather poor and additional
experimental observations are desired. In this study, two experiments are
described, both related to the problem. First, the existence of the scaling
variable \cite{Fisher}

\begin{equation}      \label{x=}
x=(B-B_c)/T^{1/y}
\end{equation}
is reexamined by analyzing magnetotransport in amorphous In$_2$O$_x$ films
\cite{InO} ($y$ is the product of two critical exponents from the theory
\cite{Fisher}). The existing studies leave some doubts. The amorphous Mo--Ge
films from \cite{Kapit} display only 5\%
increase of the resistance on ten-fold decrease in temperature in
high-magnetic-field limit and behave not like an insulator, but more like a
metal with a small quantum correction to the resistance. The measurements
with In--O films \cite{HebPaa} were made in quasireentrant region (see below,
next section). Our measurements do not reject the possibility of the
field-induced phase transition but lead to more general scaling relations.

The problem can be approached from another side, by looking for phenomena
which would reveal existence of localized pairs in insulators. To that
purpose, transport properties of the high-resistance metastable insulating
alloy Cd-Sb were studied \cite{gd} where localized pairs were supposed to
exist. This experiment is described in the last section of the paper.

\centerline {\bf In--O --- SCALING RELATIONS~~ \cite{InO}}

There are two main types for the sets of 2D field-tuned SIT $R(T)$ curves.
They differ in the behavior below the transition onset temperature $T_{c0}$
(Fig.1). The first type is more complicate. In fields weak enough, the curves
are maximum near $T_{c0}$ and below go monotonously down, with positive first
derivative, until the resistance $R$ reaches zero (if at all).  In fields
strong enough, the first derivative is negative everywhere. And there is an
intermediate field range with two extrema on the curves $R(T)$: a maximum at
$T_{\rm max}\approx T_{c0}$ and a minimum at a lower temperature.  The latter
separates from the maximum and shifts to lower $T$ with decreasing field.

The second type is simplier. There is no intermediate field range and no
curves with minima at all; the rise in the field results in the shift of the
maximum to lower $T$ untill it disappears. The authors of \cite{Gold}, basing
on their results obtained with ultrathin films of metals, consider the second
type as an ideal one and attribute the low-temperature minimum (the so-called
'quasireentrant transition') to inhomogeneities and to one-particle tunneling
between superconducting grains.

Amorphous In--O films described in \cite{HebPaa} revealed the behavior of the
first type, amorphous Mo--Ge films \cite{Kapit} -- of the second type. The
scaling variable (\ref{x=}) proved to work in both cases.  Both experiments
had one feature in common: the films had low $T_{c0}$ values, below 0.3\,K. As
$T_{c0}$ seems to be the main energy which determines the scale for the problem,
one should expect to get similar scaling relations with films of higher $T_{c0}$
in higher range of temperatures. We checked this in experiments with
amorphous In--O films 200\,\AA~thick \cite{InO}. The value of $T_{c0}$ in such
films depends on the oxygen content and hence on the heat treatment
\cite{ShOv,gg}. We encountered films of both types. In Fig.1 examples are
presented with two films brought by corresponding heat treatment to
close values of $T_{c0}$. Below we shall present a detailed analysis for a
film of the second type only since it is supposed to be more homogeneous
\cite{Gold}.

The analysis in \cite{HebPaa,Kapit} started with determining the values of the
critical field $B_c$ and of the critical resistance $R_c$ by using the
condition

\begin{equation}			\label{partial}
(\partial R/\partial T)\Bigr|_{B_c}=0
\end{equation}
at the lowest temperature and assuming that isomagnetic curve at $B_c$
remained zero-slope till $T=0$. We did not have a horizontal curve in our
fan, Fig.2.  Hence, instead, we marked maxima on the curves $R(T)\Bigr|_B$
and, utilizing linear relation between maximum values $R_{\rm max}$ and their
positions, extrapolated the function $R_{\rm max}(T)$ to $T=0$ and got the
limiting $R_c$ value (dashed line in Fig.2). One of the curves of the fan
in Fig.2, the curve obtained at $B=5\,$T looks like a straight line; it
separates curves with positive second derivative from curves with negative
derivative. Extrapolated to zero temperature (dotted line), it comes
practically to the same point $R_c$. This gives us grounds to claim that,
inspite we remain rather far in temperature from the supposed transition, at
$T>0.3\,K$, we can determine the critical parameters in the state in question
of our sample from Fig.2:

\begin{equation}			\label{parameters}
B_c=\mbox{5 T},\qquad R_c=\mbox{8 kOhm}.
\end{equation}
In essense, we have assumed that not the first but the second derivative is
zero at the transition point; the first derivative remains finite and the
line which separates on the $(T,R)$-plane two Fisher phases, if they exist,
{\it has a finite slope}. For this definite film the slope is

\begin{equation}			\label{slope}
\gamma\equiv(\partial R/\partial T)\Bigr|_{B_c}=-\mbox{0.83 kOhm/T}.
\end{equation}
Next, we eliminate this slope by introducing a function

\begin{equation}				\label{Rtilde}
\tilde R=R(T)-\gamma T
\end{equation}
and check the scaling properties not of $R$ but of $\tilde R$. Plotting
$\tilde R$ against the scaling variable $\mid B-B_c\mid/T^{1/y}$ we selected
the power $y$ to bring the data to the same curve, the critical field $B_c$
being kept within 1\,\%
of the (\ref{parameters}) value (Fig.3). The adjustment led to the same value
of $y$, $y=1.3$, which had been obtained in \cite{HebPaa} and \cite{Kapit}.
For an additional comparison, two scaling curves from \cite{Kapit} for
$R(T,B)$ normalized by $R_c=8.1\,$T are reproduced in Fig.3.

Similar analysis was applied to the data for an amorphous In--O film with the
first type of behavior (Fig.1(a)). This time we used not maxima, but minima on
the curves $R(T)\Bigr|_B$ and got the $R_c$ value by extrapolation
to $T=0$ of the function $R_{\rm min}(T)$. The main difference consisted in
the sign of the slope $\gamma$. The deformed function $\tilde R(T,B)$ after
using the scaling variable (\ref{x=}) again collapsed onto two branches and
even the value of $y$ was obtained practically the same.

These results mean that the scaling relations \cite{Fisher} either should be
generalized by replacing the film resistance $R$ by $\tilde R$ from
(\ref{Rtilde}) or they are not very specific and hence not very crucial as a
criterion of existence of the field-tuned phase transition.

\centerline{\bf Cd--Sb --- PAIR TUNNELING~~\cite{gd}}

The main process which determines the low-temperature conductance in
insulators is tunneling. Suppose that we have localized Cooper pairs. For an
electron of a pair to hop, it has to "pay off" the binding energy $\Delta$.
This introduces into conductivity additional exponential factor

\begin{equation}
\sigma_1(T) \approx \sigma_n\exp(-\Delta/T),      \label{a1}
\end{equation}
where $\sigma_n$ is the expected conductivity at a temperature $T$ in the
absence of pairing. High enough magnetic field is assumed to destroy the
pairing. The phenomenon can be recognized by negative magnetoresistance.
Field eliminates the gap and makes the resistance $\exp(\Delta/T)$ times
smaller.  In principle, measuring the ratio
$\beta(T)=\sigma_n(T)/\sigma_1(T)$ affords the possibility of extracting
the gap $\Delta$. Such negative magnetoresistance caused by one-particle
tunneling is well known in granular superconductors, both, in films
\cite{1d,2d} and bulk \cite{6d}.

We attempted to measure the function $\beta(T)$ on metastable Cd--Sb
alloy in a high-resistance state which was for sure insulating and still
had negative magnetoresistance originated in superconductivity \cite{7d}.  A
sample of Cd$_{47}$Sb$_{53}$ alloy had a rod-like form, with all dimensions
amounting to several millimeters. The sample was transformed into a metallic
phase, which is a superconductor with $T_{c0}\approx$\,4.5\,K, in a
high-pressure chamber and, being retained in this state, was quenched to the
liquid-nitrogen temperature.  After cooling, the sample was clamped in a
holder by two pairs of gold wires with pointed ends and placed into a
cryostat. The low-temperature transport measurements were alternated with
heating of the sample to room temperature. At room temperature, the material
slowly transformed into a disordered-insulator state. The transformation was
monitored by resistance measurements and could be interrupted by returning to
the liquid-nitrogen temperature. The experimental procedure is described in
detail in \cite{7d}.

The sample resistance $R$ is in essence the averaged value, because in the
course of the transformation the sample becomes inhomogeneous. According to
Ref.~\cite{8d}, below $T_{c0}$ the sample looks like a mixture of different
"weakly-superconducting" elements, such as tunnel junctions, constrictions,
thin wires, etc. The density and scales of these elements change during
transformation, leading to evolution of the sample behavior below $T_{c0}$
\cite{6d,7d}. We'll deal with the state in which the average sample
resistance $R$ has increased by five orders of magnitude comparing to the
initial one, only a weak kink has remained on the curve $R(T)$ at $T=T_{c0}$
and the curve has negative derivative everywhere. The $T_{c0}$ value while
evolution to this state has slightly decreased down to $\approx3.8$\,K.

The low temperature measurements were performed in an Oxford TLM-400 dilution
refrigerator with a base temperature of 25 mK. A four-terminal lock-in
technique at a frequency of 10~Hz was used. The ac current was equal to 1~nA
and corresponded to the linear regime. The low-temperature part of the curve
$R(T)$ is shown in Fig.4.

The experimental dependences of $\alpha(T,B)=R(T,B)/R(T,B=4\,{\rm T})$ on
magnetic field at different temperatures are presented in Fig.5. The
magnetoresistance is negative and saturates in magnetic fields exceeding
2\,T, bringing the value of $\alpha$ to unity. So, the resistance values at a
field of 4\,T, used for normalizing, are taken from the field-independent
region.  Note that $\beta(T)=\alpha(T,0)$. In the temperature range between
490 and 190 mK the value of $\beta(T)$ increases with decreasing temperature,
in qualitative agreement with Eq.~(\ref{a1}) (inset to Fig.5). This
corresponds to what has been observed previously at higher temperatures
\cite{7d}.  However, in the low-temperature limit the behavior of $\beta(T)$
changes drastically. It is maximum at a temperature of about 100\,mK and then
decreases with further lowering the temperature. In weak magnetic fields
there appears an initial increase on the dependence. However, at any fixed
magnetic field $B<2$~T the value of $\alpha$ behaves similarly (see inset to
Fig.5). This contrasts sharply with the temperature dependence of the
normal-state resistance which is monotonous in the range of temperatures used
(see Fig.4).

The observed decrease of the ratio $\beta$ with decreasing temperature
unambiguously indicates that at low temperatures the conductivity $\sigma_1$,
which originates from single-particle tunneling, is shunted by conductivity
$\sigma_2$ of another kind:

\begin{equation}
\sigma=\sigma_1(T)+\sigma_2.      \label{a3}
\end{equation}
We believe that $\sigma_2$ is due to incoherent pair tunneling (coherent,
i.e. Josephson, pair tunneling is supposed to be absent in this insulating
state; Probably the maximum of $\alpha(B)$ at $B\approx 0.1$\,T at the lowest
temperatures designates destroying by the magnetic fields of the remnants of
the coherent scattering). The single-particle tunneling current $i_1$ is
described in the first-order approximation by barrier transparency $t$:
$i_1\propto t\exp(-\Delta/T)$. It is proportional to the product of two small
factors, one of which is temperature dependent. Since the Cooper pairs are at
the Fermi level, the two electrons forming a pair do not need to be excited
above the gap for simultaneous tunneling. Hence $i_2\propto t^2$, without the
exponential temperature-dependent factor. When the temperature is
sufficiently low, so that

\begin{equation}												\label{ea}
t>\exp(-\Delta/T), \qquad \mbox{i.e.,} \qquad T<\Delta/\mid\ln t\mid,
\end{equation}
the single-particle tunneling is frozen out, and the two-particle tunneling
current comes into play.

The two particles bound into a pair in the initial state may come to be
unbound in the final state. Such process of pair tunneling looks similar to
the two-particle contribution to the tunnel current through a superconductor
-- normal-metal junction (SIN junction) \cite{Naz}. The latter may prove to
be very important in high-resistance granular superconductors.

\centerline{\bf CONCLUSION}

Both experiments described above can be interpreted as confirmation of
existence of localized pairs. But they do not give any information about how
this localization is realized. The one-particle localization radius $\xi$ may
turn to be either larger, or smaller than the coherence length $\xi_{sc}$.
The case $\xi\gg\xi_{sc}$ is an extreme limit of granular superconductors,
with only one pair in a grain. The opposite case $\xi\ll\xi_{sc}$ is assumed,
for instance, in the model of localized bipolarons \cite{alex}. To
distinguish these two possibilities, other experiments are required.

I am grateful to A.I.Larkin for illuminating discussions. This work was
supported by Grants RFFI~96-02-17497 and INTAS-RFBR~95-302 and by the
"Statistical Physics" program from the Russian Ministry of Sciences.

\begin{figure}
\caption{Temperature dependence of the resistance of two different amorphous
In--O films, both 200\,\AA\ thick, in various magnetic fields. Sets (a) and
(b) represent two types of films behavior (see text)}
\end{figure}

\begin{figure}
\caption{Central part of the set from Fig.1(b) represented in detail, with
smaller field step. Experimental points are plotted only on one curve, those
at $B=5.0\,$T. Dashed line displays maxima values, $R_{\rm max}(T)$ 
extrapolated  to $T=0$. Dotted line is linear extrapolation of data at the
field $B=5.0\,$T}

\end{figure}

\begin{figure}
\caption{Scaling dependence of the renovated resistance $\tilde R$ for the
film presented in Figs.\,1(b) and~2. Different symbols are used to
distinguish the data in different fields. For comparison, the scaling
function from Fig.3 of Ref.3 normalized by $R_c=8.1$~kOhm is reproduced by
solid lines} \end{figure}

\begin{figure}
\caption{Low-temperature dependence of the resistance of the Cd--Sb sample in
the state under investigation illustrating its insulating behavior and
negative magnetoresistance.}
\end{figure}

\begin{figure}
\caption{Normalized magnetoresistance $\alpha(T,B)$ of the Cd--Sb sample as a
function of $B$ at different temperatures.  Inset: $\alpha$ as a function of
$T$ at two values of $B$; $\alpha(T,0)\equiv\beta(T)$}
\end{figure}
\psfig{file=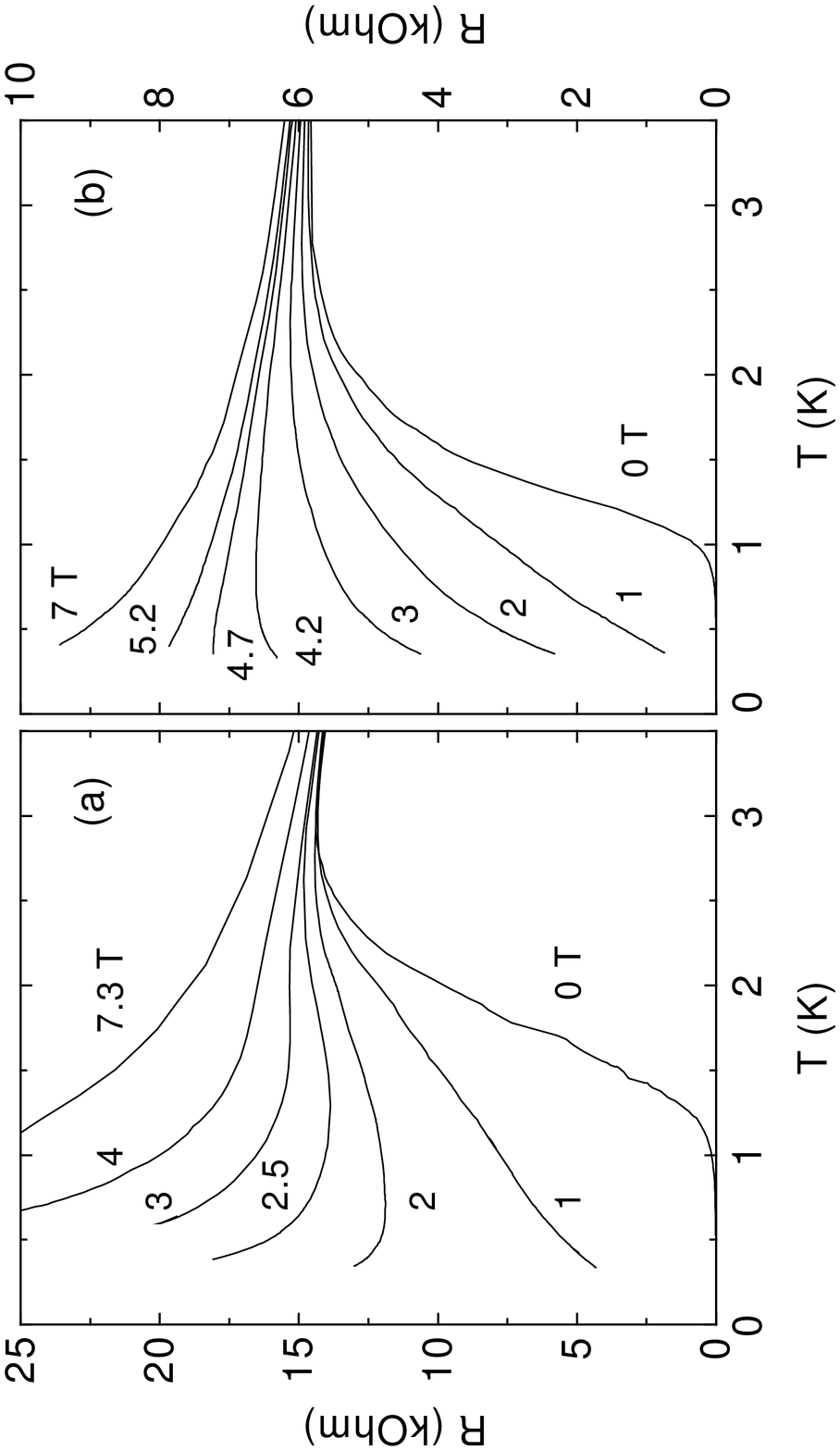}
\psfig{file=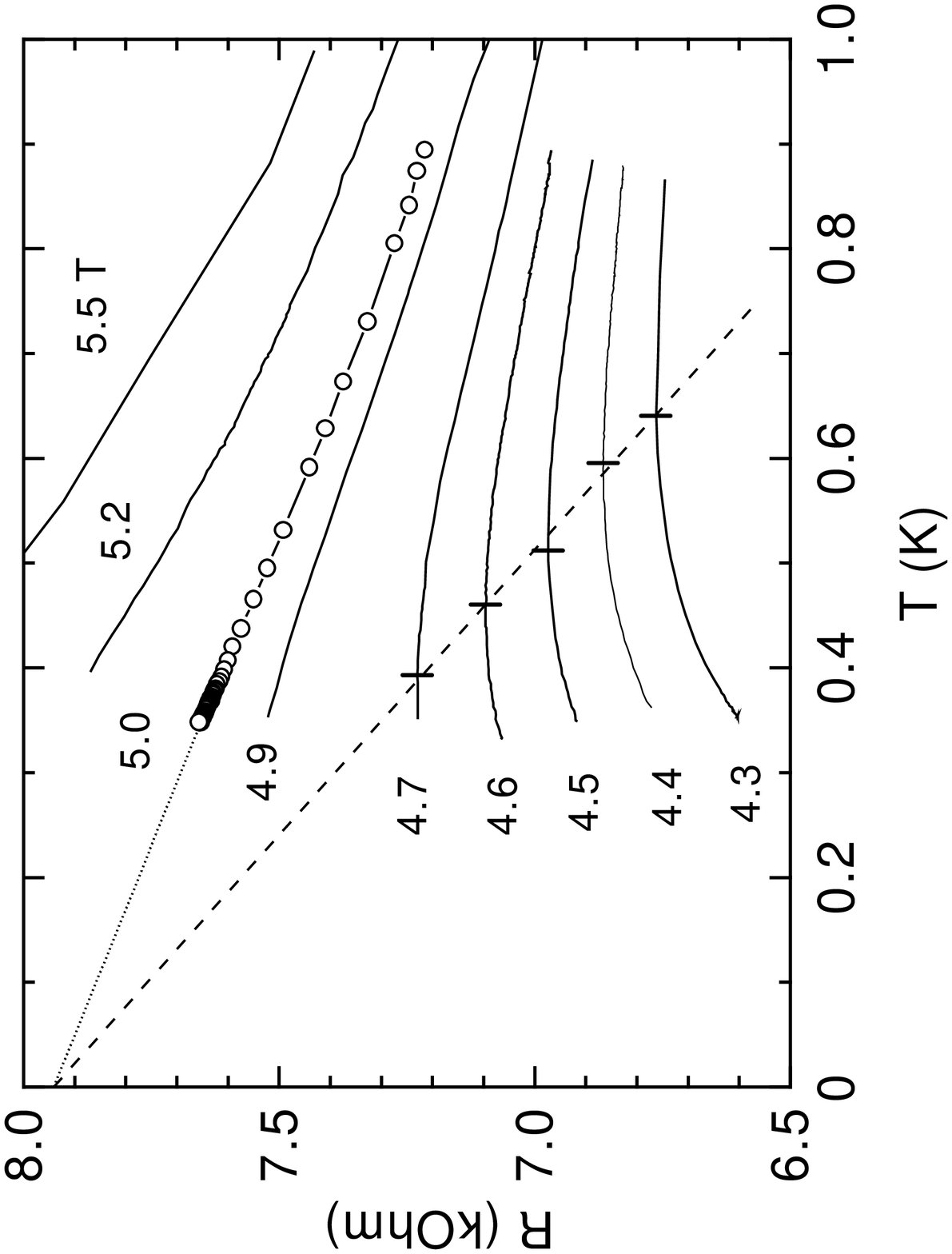}
\psfig{file=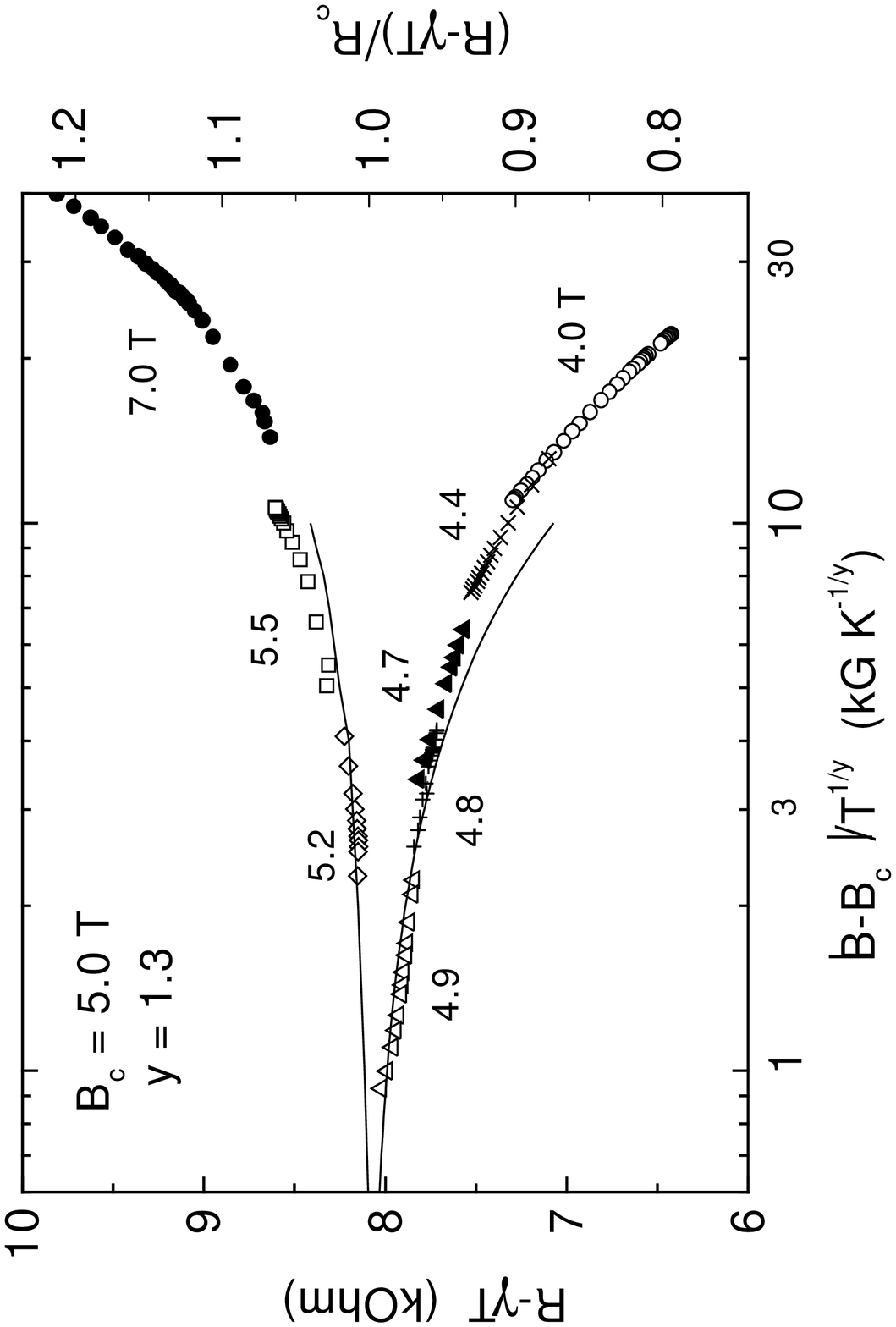}
\psfig{file=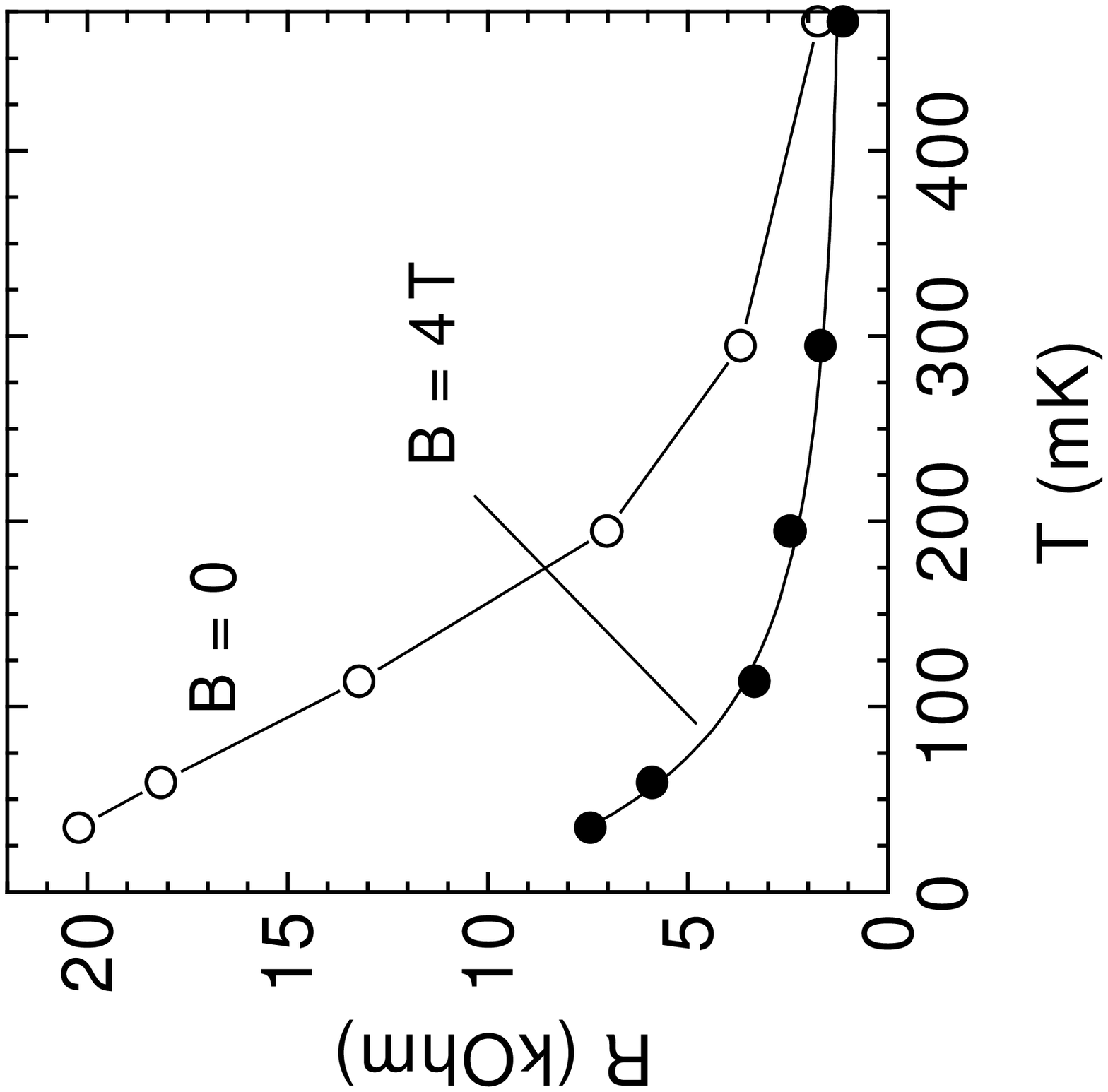}
\psfig{file=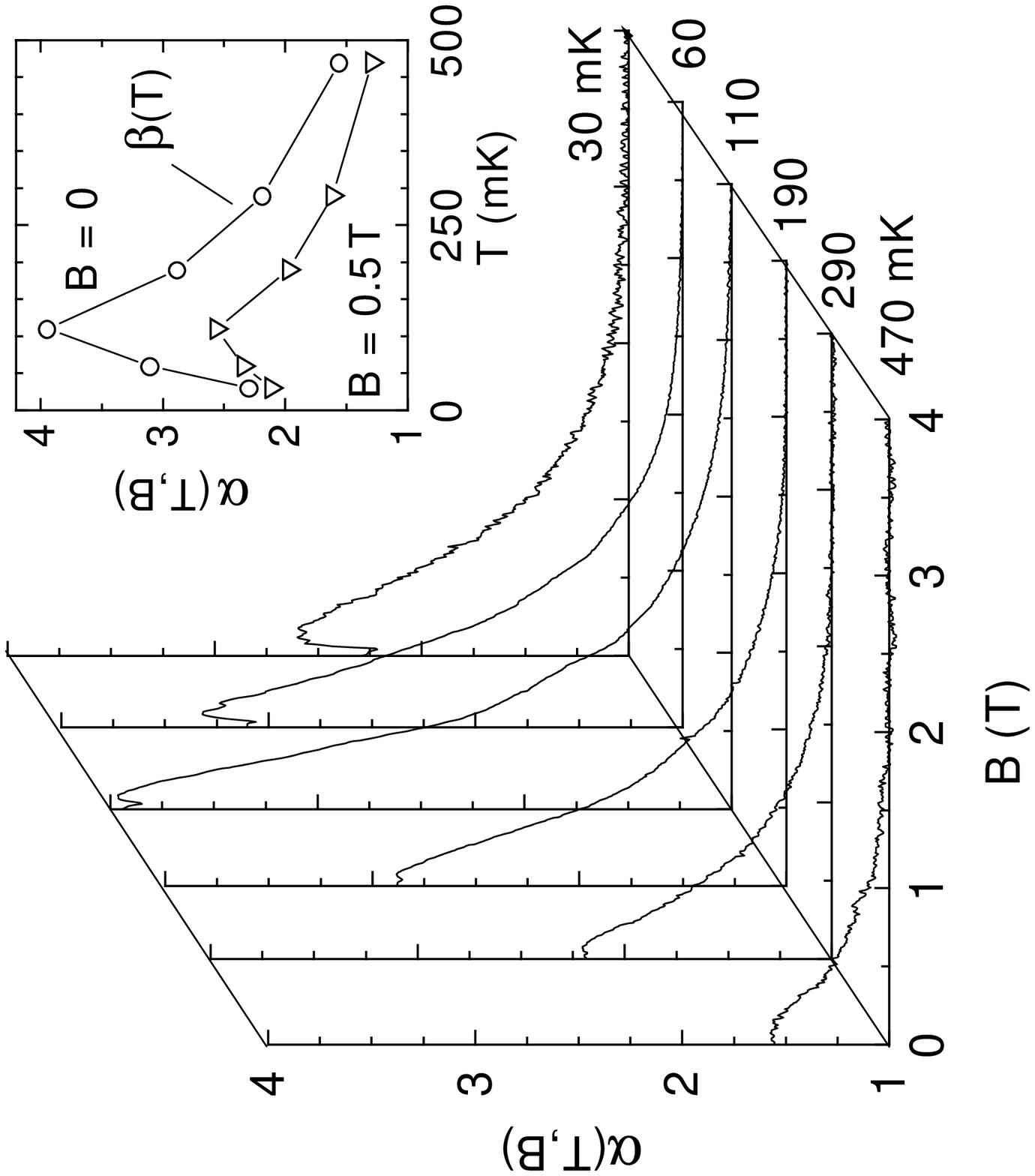}

\end{document}